# Intensity and State Estimation in Quantum Cryptography

Sindhu Chitikela

*Abstract*. This paper describes how the communicating parties can employ intensity and state estimation to detect if the eavesdropper has siphoned off and injected photons in the received communication. This is of relevance in quantum cryptography based on random rotations of photon polarizations.

**Introduction**

The paper investigates the use of state and intensity measurements by Alice and Bob to detect the presence of the active eavesdropper and the use of mixed states rather than pure states in the transmission process. This use can be applied to variations of the three-stage quantum cryptography protocols based on random unitary transformations [1]-[6].

Apart from the faked-state attack [7], [8], the main weakness of BB84 protocol is that single photons are not easy to produce, and the duplicate photons can be used by the eavesdropper to reconstruct the key. The attacker can siphon off the photons when they are transferred between Alice and Bob. Moreover, as the photons are siphoned off only at one step, the intensity of the output at the receiver's end is not affected. There is also the problem of generating single photons [9] as well as lack of availability of single photon detectors.

The polarization rotation protocols are based on fundamental cryptographic primitives [10]. The protocol called iAQC [5] is a variant of three-stage in which Alice and Bob track the intensity of the laser beam at each stage making it possible to detect the intruder. But the drawback of iAQC lies in the scenario when Eve removes some photons and replaces them with the same number of photons with different polarization back into the stream where the intensity remains constant. Such siphoning can be done by fiber tapping [11] or using half silvered mirrors if the communication media is free space. In this case the intensity is not changed and hence Eve will not be discovered and the changes in the state will be ascribed to noise. In order to overcome this threat from Eve, an additional step of detecting the state of photons using tomography [12] is considered in this paper. This new protocol is called ISA (intensity and state aware) quantum cryptography. A certain fraction of the received photons are examined for their intensity and state to determine if Eve has siphoned off photons and replaced them with other photons that can be found by determining the state.

The ISA system can be used for the standard three-stage protocol or its many variations [13]-[16] including the one-stage protocol [2]. In this paper, the process of state determination will be illustrated by examining only one transmission.

**Intensity and State Aware Protocol (ISA)**

In the ISA protocol proposed here, Alice keeps track of the state of the photons in addition to the intensity of the photons. In general, we will consider mixed states for which the density operator is a convenient representation. Assuming that a quantum system is in one of the number of states $|\psi_i\rangle$, where $i$ is an index with probabilities $p_i$, $\{p_i/\psi_i\}$ is an ensemble of pure states. The density operator of the system is defined as $\rho = \sum_i p_i |\psi_i\rangle\langle\psi_i|$. The main criterion to decide if a state is pure or mixed is considering the trace (*tr*) of the density matrix. If the *tr* is less than one then the state is said to be a mixed state else it is pure state.

Alice initially sends a bunch of photons which are in pure state and are randomly polarized to Bob. At this step, Alice computes the density matrix $\rho$ of the pure state photons. Bob rotates these photons through an angle of $\phi$ which can be either $0^o$ or $90^o$ and sends them back to Alice. As in our ping-pong protocol, if Bob chooses $\phi$ as $0^o$, it means that he has chosen bit 0. If he chooses $\phi$ as $90^o$, it means that he has chosen bit 1. Alice now computes the density matrices $\rho'$ and $\rho''$ of these photons assuming that they are rotated by $0^o$ and $90^o$ by Bob respectively.

Suppose that Eve siphons off *x* number of photons in the first stage and puts *x* number of photons with different polarization back into the stream. Therefore, a total of *2x* photons are siphoned off throughout the transmission. Eve will not be caught by the techniques that detect intruders unless the value of *2x* reaches 50% of the total number of photons sent by Alice. So, Alice has to use this technique of computation and comparison of density matrices to catch the eavesdropper. The following section illustrates the processing done at Alice and the way Eve is caught.

Figure 1 shows the ISA protocol. The process of Alice and Eve detecting the state of the photons uses quantum tomography as explained in detail in the following section.

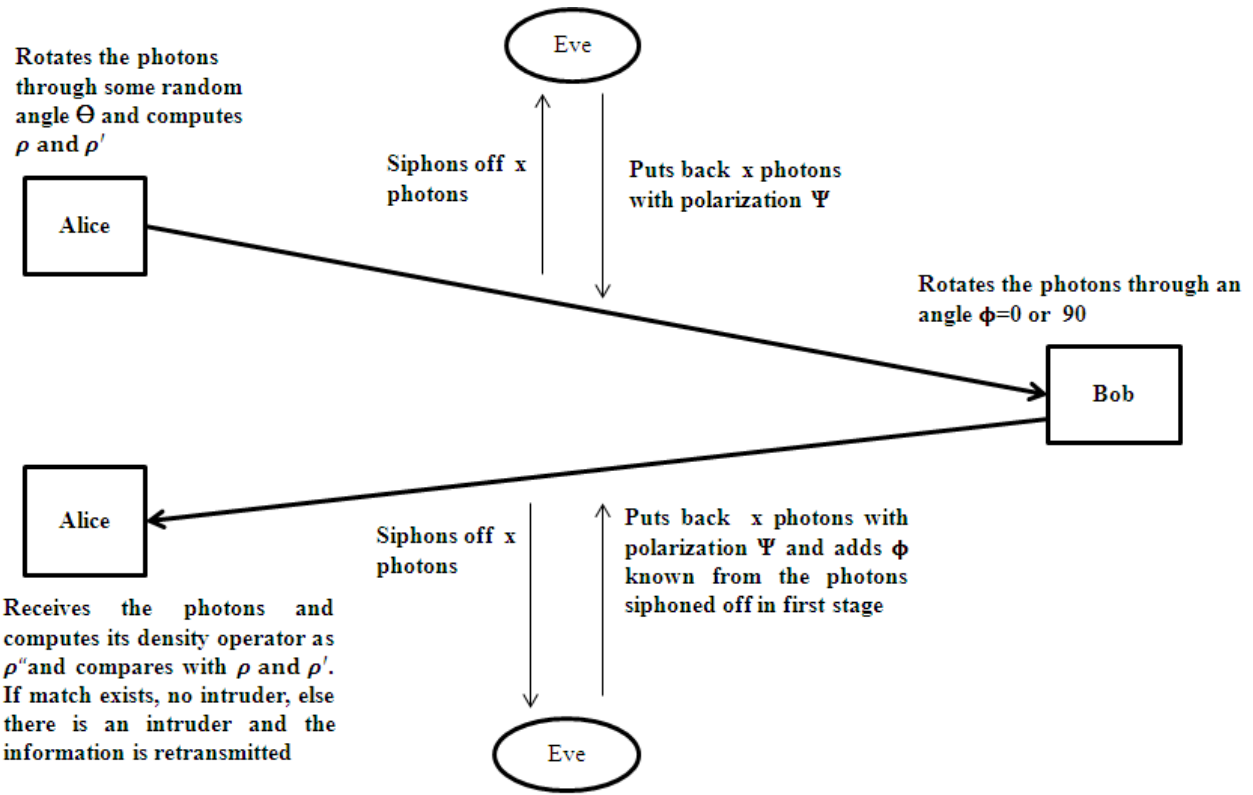

Figure 1 Intensity and State Aware Quantum Cryptography

**Tomography**

Both Alice and Eve have to measure the unknown states to the information that Bob has sent. This is achieved by quantum state tomography by which identical unknown quantum states is characterized. Initially, all the photons pairs are filtered using the spatial filters and frequency filters. After the filtering process, the unknown states are measured by the process of projection [12]. An arbitrary polarization measurement can be realized using a quarter-wave plate, a half-waveplate and a polarizing beam splitter (PBS) in an order. The quarter-wave plate and the half-waveplate are used to rotate the state to $|H\rangle$.

Then the PBS will transmit the projected state and reflect its orthogonal compliment. The resultant measurement state is sent through the detectors and a photon counting circuit is implemented to maintain the coincidental counts. Using the likelihood function, the T matrix is obtained and using the T matrix density operator can be calculated as $\rho = T^{-1}T / Tr\{T^{-1}T\}$ [12]. Once the density operator or matrix $\rho$ is obtained, any single-qubit density matrix, $\rho$ can be uniquely represented by three parameters $\{S_1, S_2, S_3\}$ are the Stokes parameters such that $\rho = \sum_{i=1}^{3}(S_i \sigma_i)$ such that $\sigma_0 = \begin{pmatrix} 1 & 0 \\ 0 & 1 \end{pmatrix}$, $\sigma_1 = \begin{pmatrix} 0 & 1 \\ 1 & 0 \end{pmatrix}$, $\sigma_2 = \begin{pmatrix} 0 & -i \\ i & 0 \end{pmatrix}$ and $\sigma_3 = \begin{pmatrix} 1 & 0 \\ 0 & -1 \end{pmatrix}$.

Measurements can also be made in any non-orthogonal bases. The following computation illustrates how Stokes parameters which correspond the measurement in the D/A (diagonal), H/V (rectilinear) and R/L (right-circular and left-circular) basis are calculated and their representation on the Poincare′ sphere is shown. Note that in our case, the Stokes parameter ($S_2$) corresponding to the R/L basis is always zero. Suppose Alice receives a bunch of photons from Bob and obtains the density matrix as $\rho = \begin{pmatrix} 0.5 & 0.5 \\ 0.5 & 0.5 \end{pmatrix}$. By the equation $S_i \equiv Tr\{\sigma_i \rho\}$, the Stokes parameters can be calculated as $S_0=1, S_1=1, S_2=0, S_3=0$. Since $\sum_{i=0}^{3} S_i = 1$, it indicates that the obtained state is pure state. Figure 3 shows the representation of these values in the Poincaré sphere. Since the state is on the surface, it is a pure state.

It is assumed that Alice picks up a random polarization angle, $\theta$ and sends pure state photons to Bob. In addition, Alice computes two density matrices, imagining that Alice makes a rotation of $0°$ in one case and $90°$ in the other case. Assuming that Bob makes a $0°$ rotation on the photons that he receives, Alice computes the density matrix, $\rho$. Assuming that Bob makes a $90°$ rotation, Alice computes the density matrix as $\rho'$.

Once Alice receives the photons from Bob in the second stage, she computes a new density matrix for that photon state as $\rho''$ and then compares it with $\rho$ and $\rho'$. If there is no match, then Alice assumes that the states that she received is not pure and she makes sure by verifying if the trace of $(\rho'')^2$ is less than one which means that it is mixed state. This is how Eve is detected in this model. Let the polarization of photons added by Eve be φ.

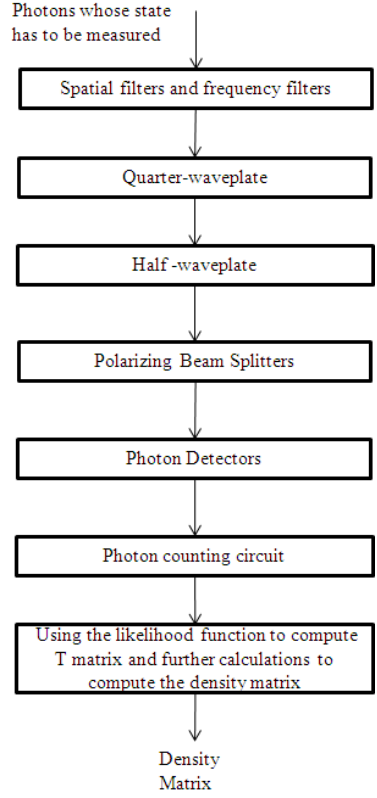

Figure 2 Quantum Tomography

The following calculations show how Eve's interference is discovered. It is assumed that Alice sends a total of 100 photons with the polarization angle $\theta = 30^\circ$. Suppose that Eve siphons off 5 photons and injects 5 other photons with a random polarization, $\varphi$ of about $45^\circ$ in the first stage. Then Bob rotates the photons by $\theta = 0^\circ$. Then in the second stage Eve takes 5 more photons and injects 5 photons with the same polarization, $\varphi$. That means 80 photons are with the state $\sqrt{3}/2|0\rangle + 1/2|1\rangle$ and 20 with state $1/\sqrt{2}|0\rangle + 1/\sqrt{2}|1\rangle$. Now the density matrices $\rho$, $\rho'$ and $\rho''$ computed by Alice are

$$\rho = \begin{bmatrix} 3/4 & \sqrt{3}/4 \\ \sqrt{3}/4 & 1/4 \end{bmatrix}; \rho' = \begin{bmatrix} 3/4 & \sqrt{3}/4 \\ \sqrt{3}/4 & 1/4 \end{bmatrix}; \rho'' = \begin{bmatrix} 0.7 & 0.4464 \\ 0.4464 & 0.3 \end{bmatrix}$$

which is quite different from that of $\rho$ and $\rho'$ and the trace of $(\rho'')^2$ is less than 1. Therefore, Eve is caught. If Alice finds that the photons are in pure state, she uses quantum tomography to find the unknown state. Any physical density matrix can be diagonalized and its corresponding eigenvalues and eigenvectors can be computed. The eigenvalues of $\rho$ are {0.0000, 1.0000} and the corresponding eigenvectors are (-0.5000, 0.8660) and (0.8660, 0.5000) respectively. The eigenvalues of $\rho''$ are {0.0108, 0.9892} and corresponding eigenvectors calculated are (-0.5437, 0.8393) and (0.8393, 0.5437) respectively. The eigenvalues of any density matrix give the intensities and their corresponding eigenvector gives the angle at each intensity.

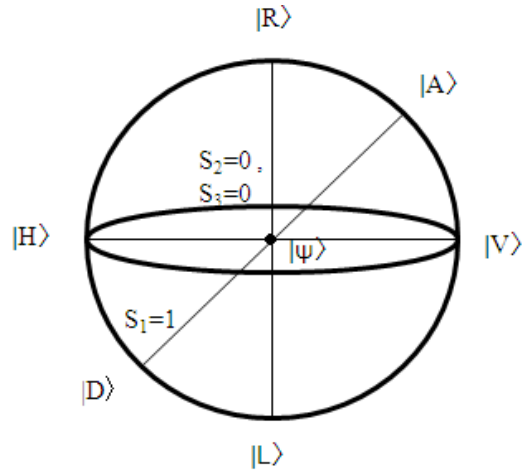

Figure 3 Poincaré sphere

The following graphs show the change in intensities as the number of photons siphoned off by Eve changes. The angle by which Alice rotates the photons is denoted by $\theta$ and the polarization of photons inserted by Eve as $\varphi$. The graph in Figure 4 indicates the intensities of photons at the highest peak obtained when $\theta = 22.5^0$ and $\varphi = 30^0$.

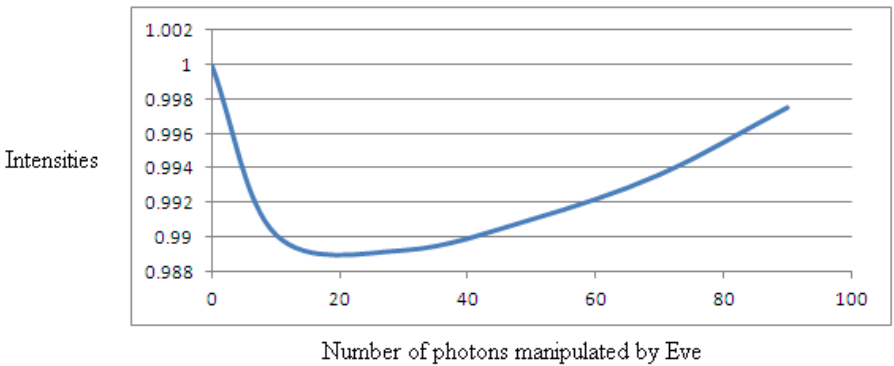

Figure 4 Graph showing the varying peak intensities with change
in number of photons manipulated by Eve when $\theta = 22.5^0$ and $\varphi = 30^0$

Figure 5 to Figure 11 indicate the angle at which the intensity is high and the peak intensities varying with the number of photons manipulated by Eve. The examples are considered with difference between angle used by Alice and Eve that is ($\theta$-$\varphi$) as $7.5^0$, $15^0$, $30^0$ and $60^0$. The intensities and the values of angle are obtained by the computation of eigenvalues and eigenvectors.

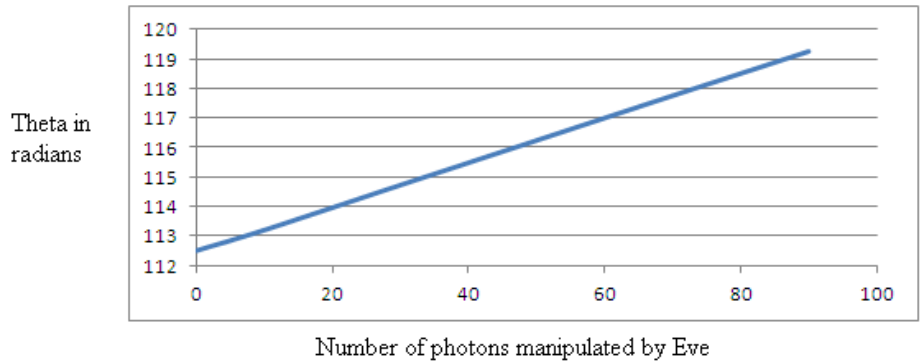

Figure 5 Graph showing the varying values of angle whose intensities are high with change in number of photons manipulated by Eve when $\theta = 22.5^0$ and $\varphi = 30^0$

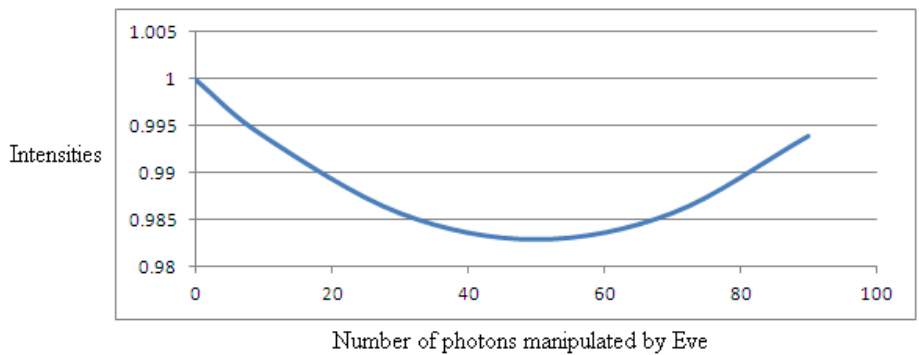

Figure 6. Graph showing the varying peak intensities with change in number of photons manipulated by Eve when $\theta = 45^0$ and $\varphi = 60^0$

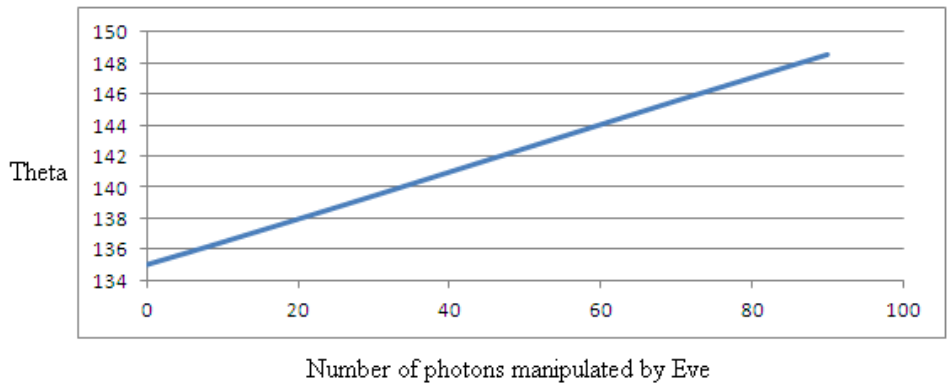

Figure 7. Graph showing the varying values of angle whose intensities are high with change in number of photons manipulated by Eve when $\theta = 45^0$ and $\varphi = 60^0$

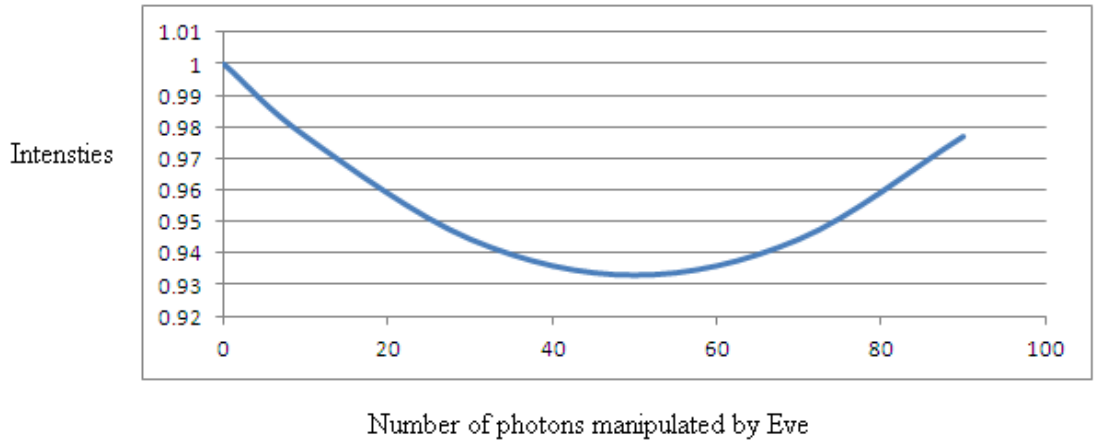

Figure 8. Graph showing the varying peak intensities with change in number of photons manipulated by Eve when $\theta = 30^0$ and $\varphi = 60^0$

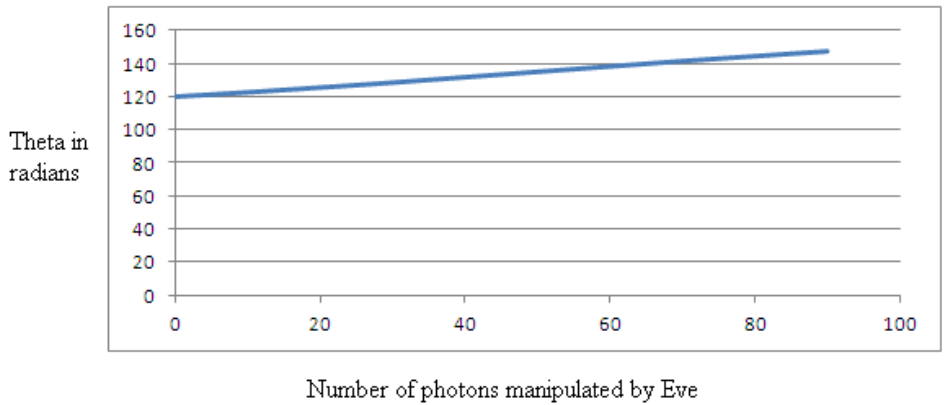

Figure 9. Graph showing the varying values of angle whose intensities are high with change in number of photons manipulated by Eve when $\theta = 30^0$ and $\varphi = 60^0$

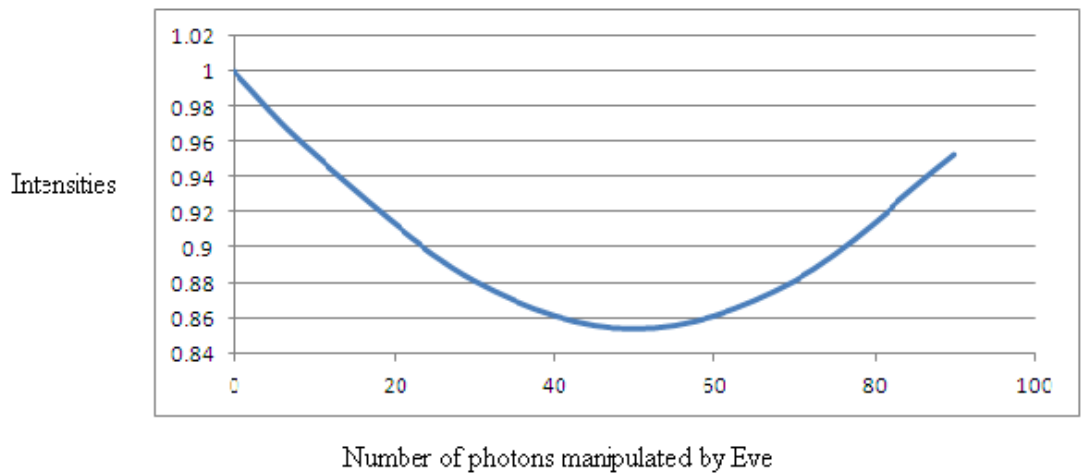

Figure 10. Graph showing the varying peak intensities with change in number of photons manipulated by Eve when $\theta = 30^0$ and $\varphi = 90^0$

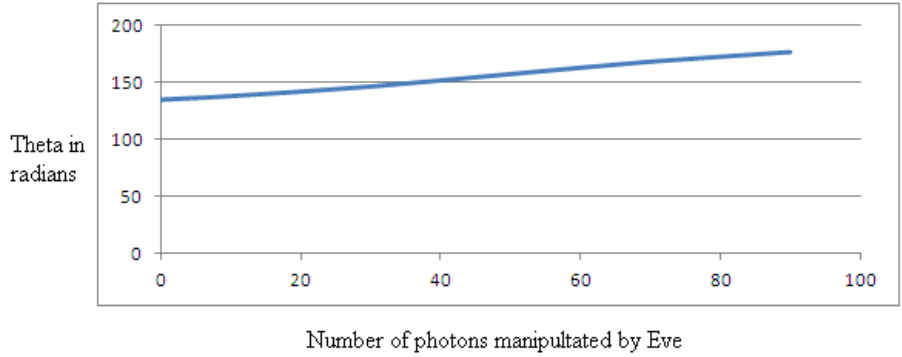

Figure 11 Graph showing the varying values of angle whose intensities are high with change in number of photons manipulated by Eve when $\theta = 30^0$ and $\varphi = 90^0$

Combining the results from all the above experiments, the graphs in Figure 12 and Figure 13 are obtained. The Figure 12 shows the varying peak intensities as the difference between $\theta$ and $\varphi$ varies. Similarly Figure 13 shows the varying angle at the peak intensities as the difference between $\theta$ and $\varphi$ varies.

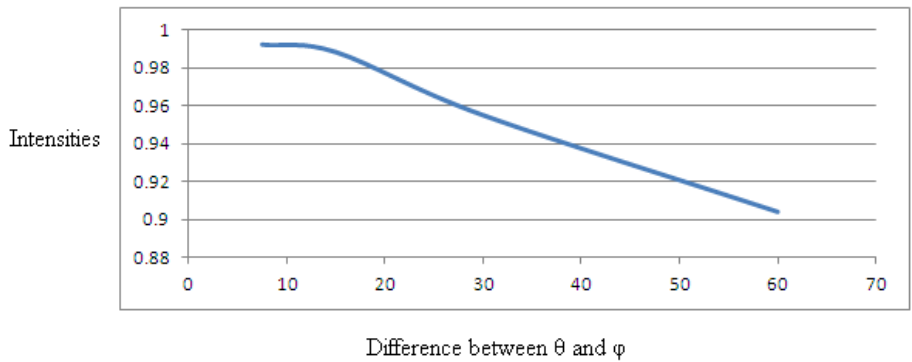

Figure 12 Graph showing the varying values peak intensities with the difference between $\theta$ and $\varphi$ equals $7.5^0$, $15^0$, $30^0$ and $60^0$

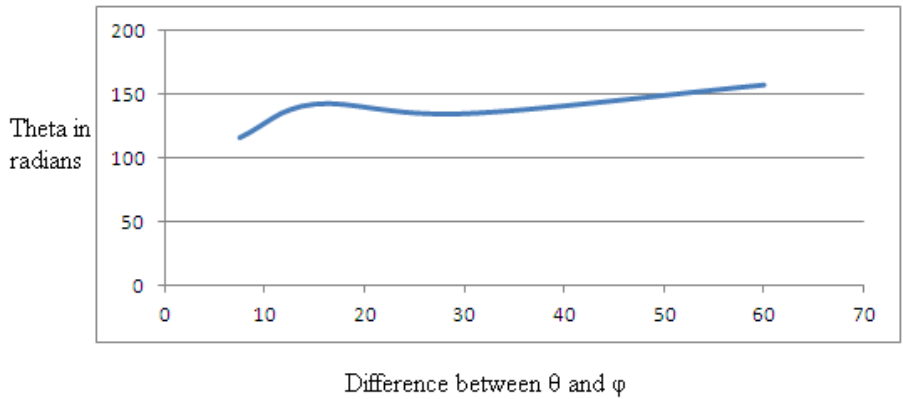

Figure 13 Graph showing the varying values of angles at peak intensities with the difference between $\theta$ and $\varphi$ equals $7.5^0$, $15^0$, $30^0$ and $60^0$

## Conclusion

The ISA protocol overcomes one of the drawbacks of iAQC and helps to detect the intruder by tracking the state of the photons. This paper has shown the influence on the main received state and angle between the polarizations chosen by Alice and those by Eve so as to determine whether Eve is present. But this has only been done in a general way. Further research needs to obtain statistical results on the confidence in the decision reached and how it is related to the number of photons that are used in the cryptographic protocol.

Further research is also needed to investigate the use of other means of coding binary information in quantum state in the communication between the two parties. This can include use of mixed quantum states. Mathematical analysis will be performed to relate the confidence in detecting the eavesdropper based on analyzed fraction of data and also the probability of false alarm. The eavesdropper's actions will be examined from a game-theoretic perspective. The performance of the various variants of the three-stage protocol will be investigated for various noise environments.